# A computational study of the aerodynamic forces and power requirements of dragonfly *Aeschna juncea* hovering

Mao Sun* and Shi Long Lan

Institute of Fluid Mechanics, Beijing University of Aeronautics & Astronautics, Beijing 100083, People′s Republic of China

Corresponding author: Mao Sun

Address: Institute of Fluid Mechanics
Beijing University of Aeronautics & Astronautics
Beijing 100083, P.R. China

*E-mail: sunmao@public.fhnet.cn.net

## Summary

Aerodynamic force generation and mechanical power requirements of a dragonfly (*Aeschna juncea*) in hovering flight are studied. The method of numerically solving the Navier-Stokes equations in moving overset grids is used.

When the midstroke angles of attack in the downstroke and the upstroke are set to $52°$ and $8°$, respectively (these values are close to those observed), the mean vertical force equals the insect weight and the mean thrust is approximately zero. There are two large vertical force peaks in one flapping cycle. One is in the first half of the cycle, which is mainly due to the hindwings in their downstroke; the other is in the second half of the cycle, which is mainly due to the forewings in their downstroke. Hovering with a large stroke plane angle ($52°$), the dragonfly uses drag as a major source for its weight supporting force (approximately 65% of the total vertical force is contributed by the drag and 35% by the lift of the wings).

The vertical force coefficient of a wing is twice as large as the quasi-steady value. The interaction between the fore- and hindwings is not very strong and is detrimental to the vertical force generation. Compared with the case of a single wing in the same motion, the interaction effect reduces the vertical forces on the fore- and hindwings by 14% and 16% of that of the corresponding single wing, respectively. The large vertical force is due to the unsteady flow effects. The mechanism of the unsteady force is that in each downstroke of the hindwing or the forewing, a new vortex ring containing downward momentum is generated, giving an upward force.

The body-mass-specific power is 37 W kg$^{-1}$, which is mainly contributed by the aerodynamic power.

**Introduction**

Dragonflies are capable of long-time hovering, fast forward flight and quick maneuver. Scientists have always been fascinated by their flight. Kinematic data such as stroke amplitude, inclination of the stroke-planes, wing beat frequency and phase-relation between the fore- and hindwings, etc., were measured by taking high-speed pictures of dragonflies in free-flight (e.g. Norberg, 1975; Wakeling and Ellington, 1997b) and tethered dragonflies (e.g. Alexander, 1984). Using these data in quasi-steady analyses (not including the interaction effects between forewing and hindwing), it was shown that the lift coefficient required for flight was much greater than the steady-state values that measured from dragonfly wings (Wakeling and Ellington, 1997a). This suggested that unsteady wing motion or/and flow interaction between the fore- and hindwings must play important roles in the flight of dragonflies (Norberg, 1975; Wakeling and Ellington, 1997c).

Force measurement on a tethered dragonfly was conducted by Somps and Luttges (1985). It was shown that over some part of a stroke cycle, lift force was many times larger than that measured from dragonfly wings under steady-state conditions. This clearly showed that the effect of unsteady flow and/or wing interaction were important. Flow visualization studies on flapping model dragonfly wings were conducted by Saharon and Luttges (1988,1989), and it was shown that constructive or destructive wing/flow interactions might occur, depending on the kinematic parameters of the flapping motion. In these studies, only the total force of the fore- and hindwings was measured, and moreover, force measurements and flow visualizations were conducted in separated works.

In order to further understand the dragonfly aerodynamics, it was desirable to have the aerodynamic force and flow structure simultaneously and also to know the force on the individual forewing and hindwing during their flapping motions. Freymuth (1990) conducted force measurement and flow visualization on an airfoil in hover modes. One of the hover modes was for hovering dragonflies. Only mean vertical force was measured. It was shown that large mean vertical force coefficient could be obtained and the force was related to a wake of vortex pairs which produced a downward jet of stream. Wang (2000) used computational fluid dynamics (CFD) method to study the aerodynamic force and vortex wake structure of an airfoil in dragonfly hovering mode. Time variation of the aerodynamic force in each flapping cycle and the vortex shedding process were obtained. It was shown that large vertical force was produce during each downstroke and the mean vertical force was enough to support the weight of a typical dragonfly. During each downstroke, a vortex pair was created. The large vertical force was explained by the downward two-dimensional jet induced by the vortex pair.

In the works of Freymuth (1990) and Wang (2000), only a single airfoil was considered. Lan and Sun (2001c) studied two airfoils in dragonfly hovering mode using CFD method. For comparison, they also computed the flow of a single airfoil. For the case of single airfoil, their results of aerodynamic force and flow structure were similar to that of Freymuth's (1990) experiment and Wang's (2000) computation. For the fore and aft airfoils flapping with 180º phase difference (counter stroking), the time variation of the aerodynamic force on each airfoil was broadly similar to that of the single airfoil; the major effect of interaction between the fore and aft airfoils was that the vertical forces on both the airfoils were decreased by approximately 20% in comparison with that of the single airfoil.

The above works (Freymuth, 1990; Wang, 2000; Lan and Sun, 2001c) which obtained aerodynamic force and flow structure simultaneously were done for airfoils. It is well known that the lift on an airplane wing of large aspect ratio can be explained by a two dimensional wing theory. But for a dragonfly wing, although its aspect ratio is relatively large, its motion is much more complex than that of an airplane wing. Three dimensional effect should be investigated. Moreover, the effect of aerodynamic interaction between the fore- and hindwings in three-dimensional case are unknown. The work of Lan and Sun (2001c) on two airfoils flapping with 180º phase difference showed that interaction between the two airfoils was detrimental to their aerodynamic performance. This result is opposite to the common expectation that wing interaction of a dragonfly would enhance its aerodynamic performance. It is of interest to investigate how interaction effect will be in the three dimensional case.

In the present study, we extend our previous two-dimensional study (Lan and Sun, 2001c) to three dimensional case. As a first step, we study the case of hovering flight. For dragonfly *Aeschna juncea* in free hovering flight, detailed kinematic data were obtained by Norberg (1975). Morphological data of the dragonfly (wing shape, wing size, wing mass distribution, weight of the insect, etc.) are also available (Norberg, 1972). On the basis of these data, the flows and aerodynamic forces and the power required for producing the forces are computed and analyzed. Because of the unique feature of the motion, i.e. the forewing and the hindwing move relative to each other, the approach of solving the flow equations over moving overset grids is employed.

## Materials and methods
### *The model wings and their kinematics*

The fore- and hindwings of the dragonfly are approximated by two flat plates. The thickness of the model wings is 1%$c$ (where $c$ is the mean chord length of the forewing). The planforms of the model wings (see Fig. 1A) are similar to those of the real wings (Norberg, 1972). The two wings have the same length but the chord length of the hindwing is larger than that of the forewing. The radius of the second moment of the forewing area is denoted by $r_2$ ($r_2 = \sqrt{\int_{S_f} r^2 \mathrm{d}S_f / S_f}$, where $r$ is radial distance and $S_f$ is the area of forewing); $r_2 = 0.61R$ ($R$ is the wing length). The flapping motions of the wings in hovering flight are sketched in Fig. 1B. The hindwing leads the forewing in phase by $180°$ (Norberg, 1975). The azimuthal rotation of the wing about the z axis (see Fig. 1C) is called "translation" and the pitching (or flip) rotation of the wing near the end of a half-stroke and at the beginning of the following half-stroke is called rotation. The speed at $r_2$ is called the translational speed.

The flapping motion of a wing is simplified as follows. The wing translates downward and upward along the stroke plane and rotates during stroke reversal (Fig. 1B). The translational velocity is denoted by $u_t$ and is given by

$$u_t^+ = 0.5\pi \sin(2\pi\tau/\tau_c + \gamma), \tag{1}$$

where the non-dimensional translational velocity $u_t^+ = u_t/U$ ($U$ is the reference velocity); the non-dimensional time $\tau = tU/c$ ($t$ is the time; $c$ is the mean chord length of the forewing, used as reference length in the present study); $\tau_c$ is the non-dimensional period of the flapping cycle; and $\gamma$ is the phase angle of the translation of the wing. The reference velocity is $U = 2\Phi n r_2$, where $\Phi$ and $n$ are the stroke amplitude and stroke frequency of the forewing, respectively. Denoting the azimuth-rotational velocity as $\dot{\phi}$, we have $\dot{\phi} = u_t/r_2$.

The angle of attack of the wing is denoted by $\alpha$. It assumes a constant value in the middle portion of a half-stroke. The constant value is denoted by $\alpha_d$ for the downstroke and by $\alpha_u$ for the upstroke. Around the stroke reversal, $\alpha$ changes with time and the angular velocity ($\dot{\alpha}$) is given by:

$$\dot{\alpha}^+ = 0.5\dot{\alpha}_0^+ \{1 - \cos[2\pi(\tau - \tau_r)/\Delta\tau_r]\}, \qquad \tau_r \leq \tau \leq \tau_r + \Delta\tau_r, \tag{2}$$

where the non-dimensional form $\dot{\alpha}^+ = \dot{\alpha}c/U$; $\dot{\alpha}_0^+$ is a constant; $\tau_r$ is the time at which the rotation starts; and $\Delta\tau_r$ is the time interval over which the rotation lasts. In the time interval of $\Delta\tau_r$, the wing rotates from $\alpha = \alpha_d$ to $\alpha = 180° - \alpha_u$. Therefore, when $\alpha_d$, $\alpha_u$ and $\Delta\tau_r$ are specified, $\dot{\alpha}_0^+$ can be determined (around the next stroke reversal, the wing would rotate from $\alpha = 180° - \alpha_u$ to $\alpha = \alpha_d$, the sign of the right-hand side of equation 2 should be reversed). The axis of the flip rotation is located at a distance of 1/4 chord length from the leading edge of the wing.

*The Navier-Stokes equations and solution method*

The Navier-Stokes equations are numerically solved using moving overset grids. For flow past a body in arbitrary motion, the governing equations can be cast in an inertial frame of reference using a general time-dependent coordinate transformation to account for the motion of the body. The non-dimensionalized three-dimensional incompressible unsteady Navier-Stokes equations, written in the inertial coordinate system *oxyz* (Fig. 1C), are as follows:

$$\frac{\partial u}{\partial x}+\frac{\partial v}{\partial y}+\frac{\partial w}{\partial z}=0, \tag{3}$$

$$\frac{\partial u}{\partial \tau}+u\frac{\partial u}{\partial x}+v\frac{\partial u}{\partial y}+w\frac{\partial u}{\partial z}=-\frac{\partial p}{\partial x}+\frac{1}{Re}(\frac{\partial^2 u}{\partial x^2}+\frac{\partial^2 u}{\partial y^2}+\frac{\partial^2 u}{\partial y^2}), \tag{4}$$

$$\frac{\partial v}{\partial \tau}+u\frac{\partial v}{\partial x}+v\frac{\partial v}{\partial y}+w\frac{\partial v}{\partial z}=-\frac{\partial p}{\partial y}+\frac{1}{Re}(\frac{\partial^2 v}{\partial x^2}+\frac{\partial^2 v}{\partial y^2}+\frac{\partial^2 v}{\partial y^2}), \tag{5}$$

$$\frac{\partial w}{\partial \tau}+u\frac{\partial w}{\partial x}+v\frac{\partial w}{\partial y}+w\frac{\partial w}{\partial z}=-\frac{\partial p}{\partial z}+\frac{1}{Re}(\frac{\partial^2 w}{\partial x^2}+\frac{\partial^2 w}{\partial y^2}+\frac{\partial^2 w}{\partial y^2}), \tag{6}$$

where *u, v* and *w* are three components of the non-dimensional velocity and *p* is the non-dimensional pressure. In the non-dimensionalization, *U, c* and *c/U* are taken as the reference velocity, length and time, respectively. *Re* denotes the Reynolds number and is defined as $Re = cU/\upsilon$, where $\upsilon$ is kinematic viscosity of the air. Equations 3 to 6 are solved using an algorithm based on the method of artificial compressibility. The algorithm was first developed by Rogers and Kwak (1990) and Rogers *et al.* (1991) for single-zone grid, and it was extended by Rogers and Pulliam (1994) to overset grids. The algorithm is outlined below.

The equations are first transformed from the Cartesian coordinate system ($x,y,z,\tau$) to the curvilinear coordinate system ($\xi,\eta,\zeta,\tau$) using a general time-dependent coordinate transformation. For a flapping wing, in order to make the transformation simple, a body-fixed coordinate system (*o'x'y'z'*) is also employed (Fig.1C). In terms of the Euler angles $\alpha$ and $\phi$ (defined in Fig.1C), the inertial coordinates (*o,x,y,z*) are related to the body-fixed coordinates (*o'x'y'z'*) through the following relationship:

$$\begin{pmatrix} x \\ y \\ z \end{pmatrix} = \begin{bmatrix} \cos\alpha\cos\phi & -\sin\phi & \cos\phi\sin\alpha \\ -\sin\alpha & 0 & \cos\alpha \\ \sin\phi\cos\alpha & \cos\phi & \sin\phi\sin\alpha \end{bmatrix} \begin{pmatrix} x' \\ y' \\ z' \end{pmatrix}. \qquad (7)$$

Using equation 7, the transformation metrics in the inertial coordinate system, $(\xi_x,\xi_y,\xi_z,\xi_\tau)$ $(\eta_x,\eta_y,\eta_z,\eta_\tau)$ and $(\zeta_x,\zeta_y,\zeta_z,\zeta_\tau)$, which are needed in the transformed Navier-Stokes equations, can be calculated from those in the body-fixed, non-inertial coordinate system, $(\xi_{x'},\xi_{y'},\xi_{z'})$, $(\eta_{x'},\eta_{y'},\eta_{z'})$ and $(\zeta_{x'},\zeta_{y'},\zeta_{z'})$, which need to be calculated only once. As a wing moves, the coordinate transformation functions vary with $(x,y,z,\tau)$ such that the grid system moves and always fits the wing. The body-fixed non-inertial frame of reference ($o'x'y'z'$) is used in the initial grid generation.

  The time derivatives of the momentum equations are differenced using a second-order, three-point backward difference formula. To solve the time discretized momentum equations for a divergence free velocity at a new time level, a pseudo-time level is introduced into the equations and a pseudo-time derivative of pressure divided by an artificial compressibility constant is introduced into the continuity equation. The resulting system of equations is iterated in pseudo-time until the pseudo-time derivative of pressure approaches zero, thus, the divergence of the velocity at the new time level approaches zero. The derivatives of the viscous fluxes in the momentum equation are approximated using second-order central differences. For the derivatives of convective fluxes, upwind differencing based on the flux-difference splitting technique is used. A third-order upwind differencing is used at the interior points and a second-order upwind differencing is used at points next to boundaries. Details of this algorithm can be found in Rogers and Kwak (1990) and Rogers et al.(1991). For the computation in the present work, the artificial compressibility constant is set to 100 (it has been shown that when the artificial compressibility constant varied between 10 and 300, the number of sub-iteration changes a little, but the final result does not change).

  With overset grids, as shown in Fig. 2, for each wing there is a body-fitted curvilinear grid, which extends a relatively short distance from the body surface, and in addition, there is a background Cartesian grid, which extends to the far-field boundary of the domain (i.e. there are three grids). The solution method for single-grid is applied to each of the three grids. The wing grids capture features such as boundary layers, separated vortices and vortex/wing interactions, etc. The background grid carries the solution to the far field. The two wing grids are overset onto the background Cartesian grid and parts of the two wing-grids overlap when the two wings move close to each other. As a result of the oversetting of the grids, there are hole regions in the wing grids and in the background grid. As the wing grids move, the holes and hole boundaries change with time. To determine the hole-fringe points, the method known as domain connectivity functions by Meakin (1993) is employed. Intergrid boundary points are the outer-boundary points of the wing grids and the hole-fringe points. Data are interpolated from one grid to another at the hole-fringe points and similarly, at the outer-boundary points of the wing grids. In the present study, the background grid does not move and the two wing-grids move in the background grid. The wing grids are generated by using a Poisson solver which is based on the work of Hilgenstock (1988). They are of O-H type grids. The background Cartesian grid is generated algebraically. Some portions of the grids are shown in Fig. 2.

For far field boundary conditions, at the inflow boundary, the velocity components are specified as freestream conditions while pressure is extrapolated from the interior; at the outflow boundary, pressure is set equal to the free-stream static pressure and the velocity is extrapolated from the interior. On the wing surfaces, impermeable wall and no-slip boundary conditions are applied, and the pressure on the boundary is obtained through the normal component of the momentum equation written in the moving coordinate system. On the plane of symmetry of the dragonfly (the $XZ$ plane; see Fig.1A), flow-symmetry conditions are applied (i.e. $w$ and the derivatives of $u$, $v$, and $p$ with respect to $y$ are set to zero).

*Evaluation of the aerodynamic forces*

The lift of a wing is the component of the aerodynamic force on the wing that is perpendicular to the translational velocity of the wing (i.e. perpendicular to the stroke plane); the drag of a wing is the component that is parallel to the translational velocity. $l_f$ and $d_f$ denote the lift and drag of the forewing, respectively; $l_h$ and $d_h$ denote the lift and drag of the hindwing, respectively. Resolving the lift and drag into the $Z$ and $X$ directions gives the vertical force and thrust of a wing. $L_f$ and $T_f$ denote the vertical force and thrust of the forewing, respectively; $L_h$ and $T_h$ denote the vertical force and thrust of the hindwing, respectively. For the forewing,

$$L_f = l_f \cos\beta + d_f \sin\phi \sin\beta, \qquad (8)$$

$$T_f = l_f \sin\beta - d_f \sin\phi \cos\beta. \qquad (9)$$

These two formulae also apply to the case of hindwing. The coefficients of $l_f$, $d_f$, $l_h$, $d_h$, $L_f$, $T_f$, $L_h$ and $T_h$ are denoted as $C_{l,f}$, $C_{d,f}$, $C_{l,h}$, $C_{d,h}$, $C_{L,f}$, $C_{T,f}$, $C_{L,h}$ and $C_{T,h}$, respectively. They are defined as

$$C_{l,f} = \frac{l_f}{0.5\rho U^2 (S_f + S_h)}, \quad \text{etc.} \qquad (10)$$

where $\rho$ is the fluid density, $S_f$ and $S_h$ are the areas of the fore- and hindwings, respectively. The total vertical force coefficient ($C_L$) and total thrust coefficient ($C_T$) of the fore- and hindwings are as follows:

$$C_L = C_{L,f} + C_{L,h}, \tag{11}$$

$$C_T = C_{T,f} + C_{T,h}. \tag{12}$$

*Data of hovering flight in* Aeshna juncea

High-speed pictures of dragonfly *Aeshna juncea* in hovering flight was taken by Norberg (1975) and the following kinematic data were obtained. For both the fore- and hingwings, the chord is almost horizontal during the downstroke (i.e. $\alpha_d \approx \beta$) and is close to the vertical during the upstroke; the stroke plane angle ($\beta$) is approximately $60°$; the stroke frequency ($n$) is 36 Hz, the stroke amplitude ($\Phi$) is $69°$; the hindwing leads the forewing in phase by $180°$. The mass of the insect ($m$) is 754 mg; forewing length is 4.74 cm; hindwing length is 4.60 cm; the mean chord lengths of the fore- and hindwings are 0.81 cm and 1.12 cm, respectively; the moment of inertial of wing-mass with respect to the fulcrum ($I$) is 4.54 g cm$^2$ for the forewing and 3.77 g cm$^2$ for the hindwing (Norberg, 1972).

On the basis of the above data, the parameters of the model wings and the wing kinematics are determined as following. The lengths of both wings ($R$) are assumed as 4.7 cm; the reference length (the mean chord length of the forewing) $c = 0.81$ cm; the reference velocity $U = 2\Phi n r_2 = 2.5$ ms$^{-1}$; the Reynolds number $Re = Uc/\upsilon \approx 1350$; the stroke period $\tau_c = U/nc = 8.58$. $\gamma$ is set as $180°$ and zero for the fore- and hindwings, respectively. Norberg (1975) did not provide the rate of wing rotation during stroke reversal. Reavis and Luttges (1988) made measurements on similar dragonflies and it was found that maximum $\dot{\alpha}$ was 10000~30000 deg./sec. Here, $\dot{\alpha}$ is set as 20000 deg./sec., giving $\dot{\alpha}_0^+ \approx 1.1$ and $\Delta\tau_r = 3.36$.

**Results and analysis**
*Test of the solver*

A single-grid solver based on the computational method described above was developed by Lan and Sun (2001a). It was tested by the analytical solutions of the boundary layer flow on a flat plate (Lan and Sun, 2001a) and by the measured unsteady forces on a flapping model fruit fly wing (Sun and Wu, 2003). A moving overset-grid solver for two dimensional flow based on the above method was developed by the same authors and it was tested by comparison with the analytical solution of the starting flow around an elliptical airfoil (Lan and Sun, 2001b,c). The two-dimensional moving overset-grids solver is extended to three-dimension in the present study. The three-dimensional moving overset-grids solver is tested here in three ways. First the flow past a starting sphere is considered, for which the approximate solution of the Navier-Stokes equations is known. Secondly, the code is tested by comparing with the results of the single-grid. Finally, the code is tested against experimental data of a flapping model fruit fly wing by Sane and Dickinson (2001).

As a first test, it is noted that in the initial stage of the starting motion of a sphere, because the boundary layer is still very thin, the flow around the sphere can be adequately treated by potential flow theory, and the flow velocity around the sphere and the drag (added-mass force) on the sphere can be obtained analytically. The acceleration of the sphere during the initial start is a cosine function of time; after the initial start, the sphere moves at constant speed ($U_s$). In the numerical calculation, the Reynolds number [based on $U_s$ and the radius ($a$) of the sphere] is set as $10^3$. Fig. 3A shows the numerical and analytical drag coefficients ($C_d$) vs. non-dimensional time ($\tau_s$) ($C_d = drag / 0.5 \rho U_s^2 \pi a^2$; $\tau_s = tU_s/2a$). Between $\tau_s = 0$ and $\tau_s \approx 0.2$, the numerical result is in very good agreement with the analytical solution. Fig. 3B shows the azimuthal velocity ($u_\theta$) at $\tau_s = 0.1$ as a function of $r/2a$ ($r$ is radial distance) with fixed azimuthal angle $\pi/2$. The numerical result agrees well with the analytical solution outside the boundary layer.

In the second test, the flow around the starting sphere is computed by the single-grid code, and the results computed using the single-grid and moving overset-grid are compared (also in Fig. 3). They are in good agreement. For the case of single-grid, the grid is of O-O type, where the numerical coordinates ($\xi, \eta, \zeta$) lie along the standard spherical coordinates. It has dimensions $100 \times 65 \times 129$. The outer boundary is set at $30a$ from the sphere. The non-dimensional time step is 0.01. Grid sizes of $68 \times 41 \times 81$ and $46 \times 27 \times 51$ were also used. By comparing the results from these three grids, it was shown that the grid size of $100 \times 65 \times 129$ was appropriate for the computation. For the case of moving overset-grids, the grid system consists two grids, one is the curvilinear grid of the sphere; another is the background Cartesian grid. The outer boundary of the sphere grid is at $1.4a$ from the sphere surface and the out boundary of the background grid is $30a$ from the sphere. The grid density is made similar to that of the single-grid.

In the third test, the set up of Sane and Dickinson (2001) is followed and the aerodynamic forces are computed for the flapping model fruit fly wing. The computed lift and drag are compared with the measured in Fig.4. In the computation, the wing grid has dimensions $109 \times 50 \times 52$ around the wing section, in normal direction and in spanwise direction, respectively; the outer boundary of the wing grid is approximately $2.0c$ from the wing. The background Cartesian grid has dimensions $90 \times 85 \times 80$ and the outer boundary is $20c$ from the wing. The non-dimensional time step is 0.02. Grid density test was conducted and it was shown that above overset grids were appropriate for the computation. In Figs 4A,B, the flapping amplitude is $60°$ and the midstroke angle of attack is $50°$; in Figs 4C,D, these quantities are $180°$ and $50°$, respectively. The magnitude and trends with variation over time of the computed lift and drag forces are in reasonably good agreement with the measured results.

*The total vertical force and thrust; comparison with insect weight*

In the calculation, the wings start the flapping motion in still air and the calculation is ended when periodicity in aerodynamic forces and flow structure is approximately reached (periodicity is approximately reached 2-3 periods after the calculation is started).

Figure 5 shows the total vertical force and thrust coefficients in one cycle, computed by two grid systems, grid-system 1 and grid-system 2. In both grid-systems, the outer boundary of the wing-grid was set at about $2c$ from the wing surface and that of the background-grid at about $40c$ from the wings. For grid-system 1, the wing grid had dimensions $29 \times 77 \times 45$ in the normal direction, around the wing and in the spanwise direction, respectively, and the background grid had dimensions $90 \times 72 \times 46$ in the *X* (horizontal), *Z* (vertical) and *Y* directions, respectively (Figure 3 shows a portion of grid-system 1). For grid-system 2, the corresponding grid dimensions were $41 \times 105 \times 63$ and $123 \times 89 \times 64$. For both grid systems, grid points of the background grid concentrated in the near field of the wings where its grid density was approximately the same as that of the outer part of the wing-grid. As seen in Fig. 5, there is almost no difference between the force coefficients calculated by the two grid-systems. Calculations were also conducted using a larger computational domain. The domain was enlarged by adding more grid points to the outside of the background grid of grid-system 2. The calculated results showed that there was no need to put the outer boundary further than that of grid-system 2. It was concluded that grid-system 1 was appropriate for the present study. The effect of time step value was considered and it was found that a numerical solution effectively independent of the time step was achieved if $\Delta\tau \leq 0.02$. Therefore, $\Delta\tau = 0.02$, was used in the present calculations.

From Fig. 5, it is seen that there are two large $C_L$ peaks in one cycle, one in the first half of the cycle (while the hindwing is in its downstroke) and the other is in the second half of the cycle (while the forewing is in its downstroke). It should be noted that by having two large $C_L$ peaks alternatively in the first and second halves of a cycle, the flight would be smoother. Averaging $C_L$ (and $C_T$) over one cycle gives the mean vertical force coefficient ($\overline{C}_L$) [and the mean thrust coefficient ($\overline{C}_T$)]; $\overline{C}_L = 1.35$ and $\overline{C}_T = 0.02$. The $\overline{C}_L$ value of 1.35 gives a vertical force of 756 mg, approximately equal to the insect weight (754 mg). The computed mean thrust (11 mg) is close to zero. That is, the force balance condition is approximately satisfied. In the calculation, the stroke plane angle, the midstroke angles of attack for the downstroke and the upstroke have been set as $\beta = 52°$, $\alpha_d = 52°$ and $\alpha_u = 8°$, respectively. These values of $\beta$, $\alpha_d$ and $\alpha_u$ give an approximately balanced flight and they are close to the observed values [$\beta \approx 60°$, during the downstroke the chord is almost horizontal (i.e. $\alpha_d \approx \beta$), and during the upstroke the chord is close to vertical].

*The forces of the forewing and the hindwing*

The total vertical force (or thrust) coefficient is the sum of vertical force (or thrust) coefficient of the fore- and hindwings. Figure 6 gives the vertical force and thrust coefficients of the fore- and hindwings. The hindwing produces a large $C_{L,h}$ peak during its downstroke (the first half of the cycle) and very small $C_{L,h}$ in its upstroke (the second half of the cycle); this is true for the forewing, but its downstroke is in the second half of the cycle. Comparing Fig. 6 with Fig. 5 shows that the hindwing in its downstroke is responsible for the large $C_L$ peak in the first half of the cycle and the forewing in its downstroke is responsible for the large $C_L$ peak in the second half of the cycle. The contributions to the mean total vertical force by the forewing and hindwing are 42% and 58%, respectively. The vertical force on the hindwing is 1.38 times of that on the forewing. Note that the area of the hindwing is 1.32 times of that of the forewing. That is, the relatively large vertical force on the hindwing is mainly due to its relatively large size.

The vertical force and thrust coefficients of a wing are the results of the lift and drag coefficients of the wing. The corresponding lift and drag coefficients $C_{l,f}$, $C_{d,f}$, $C_{l,h}$, and $C_{d,h}$ are shown in Fig. 7. For the hindwing, $C_{d,h}$ is larger than $C_{l,h}$ during the downstroke of the wing; and $\beta$ is large ($52°$). As a result, a large part of $C_{L,h}$ is from $C_{d,h}$ (approximately 65% of $C_{L,h}$ is from $C_{d,h}$ and 35% is from $C_{l,h}$). This is also true for the forewing. That is, the dragonfly uses drag as a major source for its weight supporting force when hovering with a large stroke plane angle.

*The mechanism of the large vertical force*

As shown in Fig. 6, the peak value of $C_{L,h}$ is approximately 3.0 (that of $C_{L,f}$ is approximately 2.6). Note that in the definition of the force coefficient, that total area of the fore- and hindwings ($S_f + S_h$) and the mean flapping velocity $U$ are used as reference area and reference velocity, respectively. For the hindwing, if its own area and the instantaneous velocity are used as reference area and reference velocity, respectively, the peak value of the vertical force coefficient would be $3.0 \times [(S_f + S_h)/S_h] \times U^2/(\pi U/2)^2 = 2.1$. Similarly, for the forewing, the peak value would be $2.6 \times [(S_f + S_h)/S_f] \times U^2/(\pi U/2)^2 = 2.4$. Since the thrust coefficients $C_{T,f}$ and $C_{T,h}$ are small, $C_{L,f}$ and $C_{L,h}$ can be taken as the coefficients of the resultant aerodynamic force on the fore- and hindwings, respectively. The above shows that the peak value of resultant aerodynamic force coefficient for the forewing or hindwing is $2.1 - 2.4$ (when using the area of the corresponding wing and the instantaneous velocity as reference area and reference velocity, respectively). This value is approximately twice as large as the steady-state value measured on a dragonfly wing at $Re = 730 - 1890$ (steady-state aerodynamic forces on the fore- and hindwings of dragonfly *Sympetrum sanguineum* were measured in wind-tunnel by Wakeling and Ellington, 1997a; the maximum resultant force coefficient, obtained at angle of attack of around $60°$, was approximately 1.3).

There are two possible reasons for the large vertical force coefficients of the flapping wings: one is the unsteady flow effect; the other is the effect of interaction between the fore- and hindwings (in steady-state wind-tunnel test, interaction between fore- and hindwings was not considered).

*The effect of interaction between the fore- and hindwings*

In order to investigate the interference effect between the fore- and hindwings, we computed the flow around a single forewing (and also a single hindwing) performing the same flapping motion as above. Figure 8A,B gives vertical force ($C_{L,sf}$) and thrust ($C_{T,sf}$) coefficients of the single forewing, compared with $C_{L,f}$ and $C_{T,f}$, respectively. The differences between $C_{L,sf}$ and $C_{L,f}$ and between $C_{T,sf}$ and $C_{T,f}$ show the interaction effect. Similar comparison for the hindwing is given in Fig.

8C,D. For both the fore- and hindwings, the vertical force coefficient on single wing (i.e. without interaction) is a little larger than that with interaction. For the forewing, the interaction effect reduces the mean vertical force coefficient by 14% of that of the single wing; for the hindwing, the reduction is 16% of that of the single wing. The interaction effect is not very large and is detrimental to the vertical force generation.

*The unsteady flow effect*

The above results show that the interaction effect between the fore- and hindwings is small and, moreover, is detrimental to the vertical force generation. Therefore, the large vertical force coefficients produced by the wings must be due to the unsteady flow effect. Here the flow information is used to explain the unsteady aerodynamic force.

First, the case of single wing is considered. Figure 9 gives the iso-vorticity surface plots at various times during one cycle. In order to correlate force and flow information, we express time during a stroke cycle as a non-dimensional parameter, $\hat{t}$, such that $\hat{t} = 0$ at the start of the cycle and $\hat{t} = 1$ at the end of the cycle. After the downstroke of the hindwing has just started ($\hat{t} = 0.125$; Fig. 9A), a starting vortex is generated near the trailing edge of the wing and a leading edge vortex (LEV) is generated at the leading edge of the wing; the LEV and the starting vortex are connected by the tip vortices, forming a vortex ring. Through the downstroke (Fig. 9B,C), the vortex ring grows in size and moves downward. At stroke reversal (between $\hat{t} \approx 0.36$ and $\hat{t} \approx 0.65$), the wing rotates and the LEV is shed. During the upstroke, the wing almost does not produce any vorticity. The vortex ring produced during the downstroke is left below the stroke plane (Fig. 9D,E,F) and will convect downwards due to its self-induced velocity. The vortex ring contains a downward jet (see below). We thus see that in each cycle, a new vortex ring carrying downward momentum is produced, resulting in an upward force. This qualitatively explains the unsteady vertical force production. Figure 10 gives the velocity vectors projected in a vertical plane that is parallel to and 0.6$R$ from the plane of symmetry of the insect. The downward jet is clearly seen.

Figure 11 gives the iso-vorticity surface plots for the fore- and hindwings (in the first half of the cycle the hindwing is in its downstroke; in the second half of the cycle the forewing in its downstroke). Similar to the case of single wing, just after the start of the first half of the cycle, a new vortex ring is produced by the hindwing (Fig. 11A); this vortex ring grows in size and convects downwards (Fig. 11A,B,C). Similarly, just after the start of the second half of the cycle, a new vortex ring is produced by the forewing (Fig. 11D), which also grows in size and convects downwards as time increasing. Figure12 gives the corresponding velocity vector plots. The qualitative explanation of the large unsteady forces on the fore- and hindwings is similar to that for the single wing.

On the basis of the above analysis of the aerodynamic force mechanism, we give a preliminary explanation for why the forewing-hindwing interaction is not strong and is detrimental. The new vortex ring, which is responsible for the large aerodynamic force on a wing, is generated by the rapid unsteady motion of the wing at a large angle of attack. As a result, the effect of the wake of the other wing is relatively small. Moreover, the wake of the other wing produces downwash velocity, resulting in the detrimental effects.

*Power requirements*

As shown above, the computed lift is enough to support the insect weight and the horizontal force is approximately zero; i.e. the force balance conditions of hovering are satisfied. Here we calculate the mechanical power output of the dragonfly. The mechanical power includes the aerodynamic power (work done against the aerodynamic torques) and the inertial power (work done against the torques due to accelerating the wing-mass).

As expressed in equation 20 of Sun and Tang (2002), the aerodynamic power consists of two parts, one due to the aerodynamic torque for translation and the other to the aerodynamic torque for rotation. The coefficients of these two torques (denoted by $C_{Q,a,t}$ and $C_{Q,a,r}$, respectively) are defined as

$$C_{Q,a,t} = \frac{Q_{a,t}}{0.5\rho U^2 (S_f + S_h)c}, \tag{13}$$

$$C_{Q,a,r} = \frac{Q_{a,r}}{0.5\rho U^2 (S_f + S_h)c}, \tag{14}$$

where $Q_{a,t}$ and $Q_{a,r}$ are the aerodynamic torques around the axis of azimuthal rotation ($z'$ axis) and the axis of pitching rotation, respectively. $C_{Q,a,t}$ and $C_{Q,a,r}$ are shown in Fig.13A,B. It is seen that $C_{Q,a,t}$ is much larger than $C_{Q,a,r}$.

The inertial power also consists of two parts (see equation 35 of Sun and Tang, 2002), one due to the inertial torque for translation and the other to the inertial torque for rotation. The coefficient of inertial torque for translation ($C_{Q,i,t}$) is defined as

$$C_{Q,i,t} = \frac{I}{0.5\rho(S_f + S_h)c^3}\ddot{\phi}^+, \tag{15}$$

where $\ddot{\phi}^+$ is the non-dimensional angular acceleration of wing translation. $C_{Q,i,t}$ is shown in Fig.13C. The inertial torque for rotation can not be calculated since the moment of inertial of wing-mass with respect to the axis of flip rotation is not available. Because most of the wing-mass is located near the axis of flip rotation, it is expected that the inertial torque for rotation is much smaller than that for translation. That is, both the aerodynamic and inertial torques for rotation might be much smaller than those for translation. In the present study, the aerodynamic and inertial torques for rotation are neglected in the power calculation.

The power coefficient ($C_p$), i.e. power non-dimensionalized by $0.5\rho U^3 (S_f + S_h)$, is

$$C_p = C_{p,a} + C_{p,i}, \tag{16}$$

where

$$C_{p,a} = C_{Q,a,t} \dot{\phi}^+, \tag{17}$$

$$C_{p,i} = C_{Q,i,t} \dot{\phi}^+. \tag{18}$$

$C_p$ of the fore- and hindwings are shown in Fig.14. In the figure, contributions to $C_p$ by the aerodynamic and inertial torques (represented by $C_{p,a}$ and $C_{p,i}$, respectively) are also shown. For the forewing (Fig.14A), the time course of $C_p$ is similar to that of $C_{p,a}$ in the downstroke and to that of $C_{p,i}$ in the upstroke; i.e. the aerodynamic power dominates over the downtroke and the inertial power dominates over the upstroke. This is also true for the hindwing (Fig. 14B).

Integrating $C_p$ over the part of a wingbeat cycle where it is positive gives the coefficient of positive work ($C_W^+$) for translation. Integrating $C_p$ over the part of the cycle where it is negative gives the coefficient of 'negative' work ($C_W^-$) for 'braking' the wing in this part of the cycle. $C_W^+$ and $C_W^-$ for the forewing are 8.33 and -2.16, respectively. For the hindwing, they are 8.93 and -1.14, respectively.

The mass specific power ($P^*$) is defined as the mean mechanical power over a flapping cycle divided by the mass of the insect, and it can be written as follows (Sun and Tang, 2002):

$$P^* = 0.5\rho U^3 (2S_f + 2S_h)(C_{W,f}/\tau_c + C_{W,h}/\tau_c)/m, \tag{19}$$

where $C_{W,f}$ and $C_{W,h}$ are the coefficients of work per cycle for the fore- and hindwings, respectively. When calculating $C_{W,f}$ or $C_{W,h}$, one needs to consider how the negative work fits into the power budget. There are three possibilities (Weis-Fogh, 1972; Ellington, 1984). One is that the negative power is simply dissipated as heat and sound by some form of an end stop, then it can be ignored in the power budget. The second is that in the period of negative work, the excess energy can be stored by an elastic element, and this energy can then be released when the wing does positive work. The third is that the flight muscles do negative work (i.e. they are stretched while developing tension, instead of contracting as in "positive" work) but the negative work uses much less metabolic energy than an equivalent amount of positive work, and again, the negative power can be ignored in the power budget. That is, out of these three possibilities, two ways of computing $C_{W,f}$ or $C_{W,h}$ can be taken. One is neglecting the negative work, i.e.:

$$C_{W,f} = (C_W^+)_{\text{forewing}}, \qquad (20)$$

$$C_{W,h} = (C_W^+)_{\text{hindwing}}. \qquad (21)$$

The other is assuming the negative work can be stored and released when the wing does positive work, i.e.:

$$C_{W,f} = (C_W^+ + C_W^-)_{\text{forewing}}, \qquad (22)$$

$$C_{W,h} = (C_W^+ + C_W^-)_{\text{hindwing}}. \qquad (23)$$

Here equations 20 and 21 are used, the computed $P^*$ is 37 W kg$^{-1}$ (when equations 22 and 23 are used, $P^*$ is 30 W kg$^{-1}$).

## Discussion

*Comparison with previous two-dimensional results*

Wang (2000) and Lan and Sun (2001c) have presented two-dimensional (2D) computations based on wing kinematics similar to those used in this study. Wang (2000) investigated the case of a single airfoil; Lan and Sun (2001c) investigated both the cases of a single airfoil and fore and aft airfoils. It is of interest to make comparison between the present three-dimensional (3D) and the previous 2D results.

The $\overline{C}_L$ value (single airfoil) computed by Wang (2000) is approximately is 1.97 [in figure 4 of Wang (2000), maximum of $u_t$ is used as reference velocity and the $\overline{C}_L$ value is approximately 0.8; if the mean of $u_t$ is used as reference velocity, the $\overline{C}_L$ value becomes $0.8 \times (0.5\pi)^2 = 1.97$]; approximately the same $\overline{C}_L$ value (single airfoil) was obtained by Lan and Sun (2001c). In the present study, the $\overline{C}_L$ values for the single forewing and single hindwing are 1.51 and 1.64, respectively, approximately 20% less than the 2D value. This shows that the 3D effect on $\overline{C}_L$ is significant. The wing length-to-chord ratio is not small (approximately 5); one might expect a small 3D effect. But for a flapping wing (especially in hover mode), the relative velocity varies along the wing span, from zero at the wing base to its maximum at the wing tip, which can increase the 3D effect. Note that although $\overline{C}_L$ is reduced by 3D effect significantly, the time course of $C_L$ of the forewing or the hindwing is nearly identical to that of the airfoil (compare Fig. 6A with figure 3 of Wang, 2000).

Lan and Sun's (2001c) results for the fore and aft airfoils showed that the interaction effect decreased the vertical forces on the airfoils by approximately 22% compared to that of the single airfoil. For the fore- and the hindwings in present study, the reduction is approximately 15%, showing that 3D forewing-hindwing interaction is weaker than the 2D case.

*Aerodynamic force mechanism and forewing-hindwing interaction*

Recent studies (e.g. Ellington et al, 1996; Dickinson et al, 1997; Wu and Sun, 2004) have shown that the large unsteady aerodynamic forces on flapping model insect wings are mainly due to the attachment of a LEV or the delayed stall mechanism. This also true for the fore- and hindwings in the present study. The LEV dose not shed before the end of the downstroke of the fore- or hindwing (Fig. 11). If the LEV sheds shortly after the start of the downstroke, the LEV would be very close to the starting vortex and a vortex ring that carries a large downward momentum (i.e. the large aerodynamic forces) could not be produced. Generation of a vortex ring carrying large downward momentum is equivalent to the delayed stall mechanism.

Data presented in Fig. 8 show that the forewing-hindwing interaction is not very strong and is detrimental. In obtaining these data, the wing kinematics observed for a dragonfly in hovering flight (e.g. 180° phase difference between the forewing and the hindwing; no incoming free-stream) has been used. Although some preliminary explanation have been given for this result, at the present, we cannot distinguish whether or not this result will exist when the phasing, the incoming flow condition, etc., are varied. Analysis based on flow simulations in which the wing kinematics and the flight velocity are systematically varied is needed.

*Power requirements compared with quasi-steady results and with* Drosophila *results*

Wakeling and Ellington (1997b,c) computed the power requirements for the dragonfly *Sympetrum sanguineum*. In most cases they investigated, the dragonfly was in accelerating and/or climbing flight. Only one case is close to hovering (flight SSan 5.2); in this case, the flight speed is rather low (advance ration is approximately 0.1) and the resultant aerodynamic force is close to the insect weight (see figure 7D of Wakeling and Ellington, 1997b and figure 5 of Wakeling and Ellington, 1997c). Their computed body-mass specific aerodynamic power is 17.1 W kg$^{-1}$ (see table 3 of Wakeling and Ellington, 1997c; note that we have converted the muscle specific power given in the table to the body-mass specific power), only approximately half the value calculated in the present study. Lehman and Dickinson (1997) and Sun and Tang (2002), based on experimental and CFD studies, respectively, showed that for fruit flies, calculation by quasi-steady analysis might under-estimate the aerodynamic power by 50%. Similar result is seen for the hovering dragonflies.

It is of interest to note that the value of $P^*$ for the dragonfly in the present study (37 W kg$^{-1}$) is not very different from that computed for a fruit fly (30 W kg$^{-1}$; Sun and Tang, 2002), even their sizes are greatly different (the wing length of the fruit fly is 0.3 cm and that of the dragonfly is 4.7 cm). For the fruit fly, the mechanical power is mainly contributed by aerodynamic power (Sun and Tang, 2002). It is approximately the case with the dragonfly in the present study (see Fig.14). From equation 15 of Sun and Tang (2002), the aerodynamic torque of a wing can be written as

$$Q_{a,t} \sim \hat{r}_d \bar{d} R, \quad (24)$$

where $\bar{d}$ is the mean drag of the wing; $\hat{r}_d$ is the radius of the first moment of the drag normalized by $R$. When the majority of the power is due to aerodynamic torque, $P^*$ can be approximated as

$$P^* \sim \hat{r}_d n \Phi R \bar{d} / \bar{L}, \quad (25)$$

$\bar{d}/\bar{L}$ is the ratio of the mean drag to the mean vertical force of the wing. For the fruit fly, this ratio is around 1 (Sun and Tang, 2002). For the dragonfly in this study, since a large part of the vertical force is contributed by the drag, this ratio is not very different from 1. We assume $\hat{r}_d$ for the two insects is not very different. Then, $P^*$ depends mainly on $n\Phi R$ (half the mean tip speed). The dragonfly's $R$ is approximately 16 times of that of the fruit fly; but its $n\Phi$ ($36\text{Hz} \times 69°$) is approximately 1/14 of that of the fruit fly ($240\text{Hz} \times 150°$). This explains why $P^*$ of the dragonfly is not very different from that of the fruit fly.

We thank the two referees whose thoughtful questions and valuable suggestions greatly improved the quality of the paper. This research was supported by the National Natural Science Foundation of China (10232010).

## Figure Legends

Fig. 1   Sketches of the model wings, the flapping motion and the reference frames. FW and HW denote fore- and hindwings, respectively. *OXYZ* is an inertial frame, with the *X* and *Y* axes in the horizontal plane; *oxyz* is another inertial frame, with the *x* and *y* axes in the stroke plane; $o'x'y'z'$ is a frame fixed on the wing, with the $x'$ axis along the wing chord and $y'$ axis along the wing span. $\beta$, stroke plane angle; $\phi$, positional angle; $\alpha$, angle of attack; *R*, wing length.

Fig. 2   Some portions of the moving overset grids.

Fig. 3   Comparison between numerical and analytical solutions of a starting sphere. (A) Drag coefficient ($C_d$) vs. non-dimensional time ($\tau_s$). (B) Azimuthal velocity ($u_\theta$) vs. non-dimensional radial distance ($r/2a$).

Fig. 4   Comparison of the calculated and measured lift and drag forces. The experimental data are reproduced from figs 3C, D of Sane and Dickinson (2001). (A, B), the midstroke angle of attack is $50°$ and stroke amplitude is $60°$; (C, D), the midstroke angle of attack is 50° and stroke amplitude is 180°.

Fig. 5   (A) Non-dimensional angular velocity of flip rotation ($\dot{\alpha}^+$) and azimuthal rotation ($\dot{\phi}^+$) of hindwing and (B) forewing; (C) time courses of total vertical force coefficient ($C_L$) and (D) total thrust coefficient ($C_T$) in one cycle.

Fig. 6   (A) Time courses of vertical force coefficients of forewing ($C_{L,f}$) and hindwing ($C_{L,h}$) and (B) thrust coefficients of the forewing ($C_{T,f}$) and the hindwing ($C_{T,h}$) in one cycle.

Fig. 7   (A) Time courses of lift coefficients of forewing ($C_{l,f}$) and hindwing ($C_{l,h}$) and (B) drag coefficients of the forewing ($C_{d,f}$) and the hindwing ($C_{d,h}$) in one cycle.

Fig. 8   (A) Time courses of vertical force coefficients of forewing ($C_{L,f}$) and single forewing ($C_{L,sf}$); (B) thrust coefficients of the forewing ($C_{T,f}$) and single forewing ($C_{T,sf}$); (C) vertical force coefficients of hindwing ($C_{L,h}$) and single hindwing ($C_{L,sh}$) and (D) thrust coefficients of the hindwing ($C_{T,h}$) and single hindwing ($C_{T,sh}$) in one cycle.

Fig. 9   Iso-vorticity surface plots at various times in one cycle (single hindwing). Note that the X axis is along the body of the dragonfly and XZ plane is the plane of symmetry of the insect. $\hat{t}$, non-dimensional time. The magnitude of the non-dimensional vorticity is 1.

Fig. 10   Velocity vectors in a vertical plane parallel to and $0.6R$ from the plane of symmetry at various times in one cycle (single hindwing). The horizontal arrow at the top left represents the reference velocity ($U$). $\hat{t}$, non-dimensional time.

Fig. 11   Iso-vorticity surface plots at various times in one cycle (fore- and hindwings). Note that the X axis is along the body of the dragonfly and XZ plane is the plane of symmetry of the insect. $\hat{t}$, non-dimensional time. The magnitude of the non-dimensional vorticity is 1.

Fig. 12   Velocity vectors in a vertical plane parallel to and $0.6R$ from the plane of symmetry at various times in one cycle (fore- and hindwings). The horizontal arrow at the top left represents the reference velocity ($U$). $\hat{t}$, non-dimensional time.

Fig. 13  (A) Time courses of aerodynamic torque coefficients for translation ($C_{Q,a,t}$) and rotation ($C_{Q,a,r}$) of forewing and (B) hindwing in one cycle; (C) time courses of inertial torque coefficient for translation ($C_{Q,i,t}$) in one cycle.

Fig. 14  Time courses of power coefficients of forewing (A) and hindwing (B) in one cycle. $C_p$, power coefficient; $C_{p,a}$, coefficient of power due to aerodynamic force; $C_{p,i}$, coefficient of power due to inertial force.

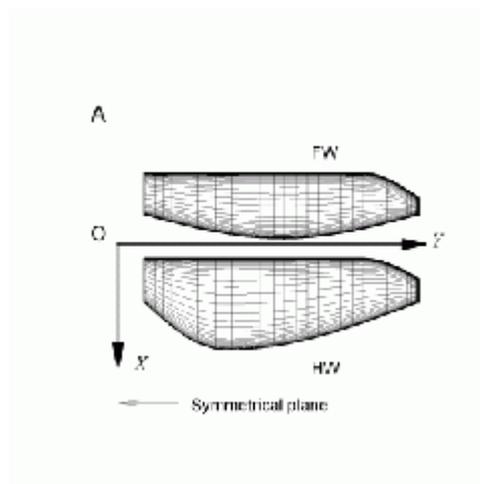

Fig.1A

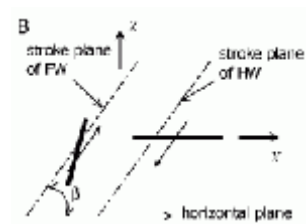

Fig.1B

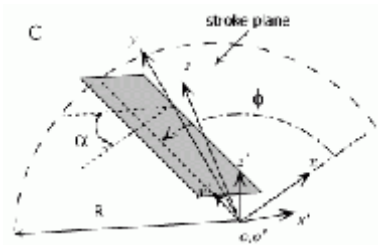

Fig.1C

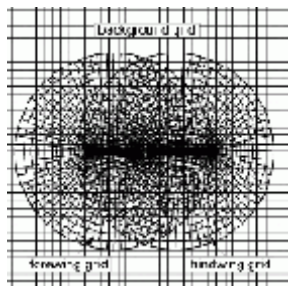

Fig.2

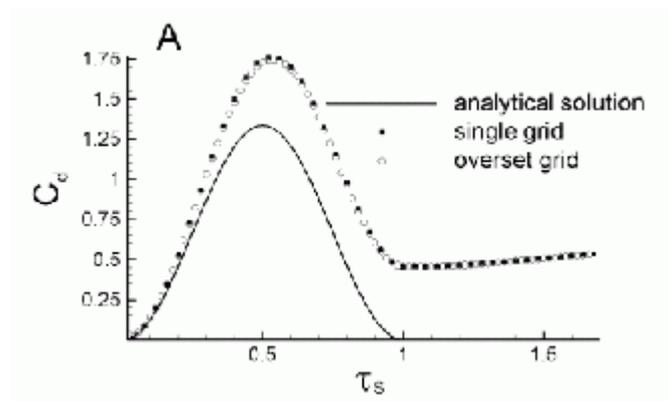

Fig.3A

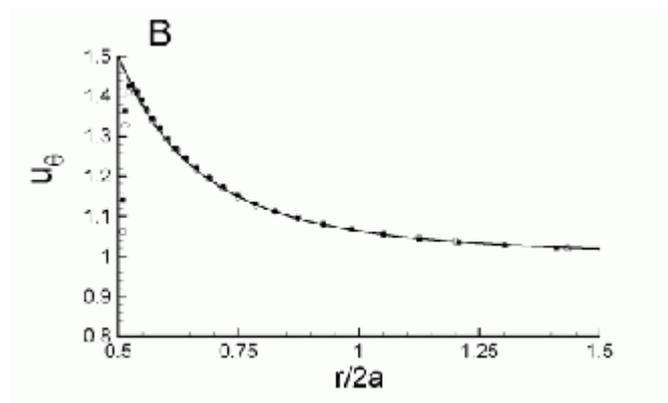

Fig.3B

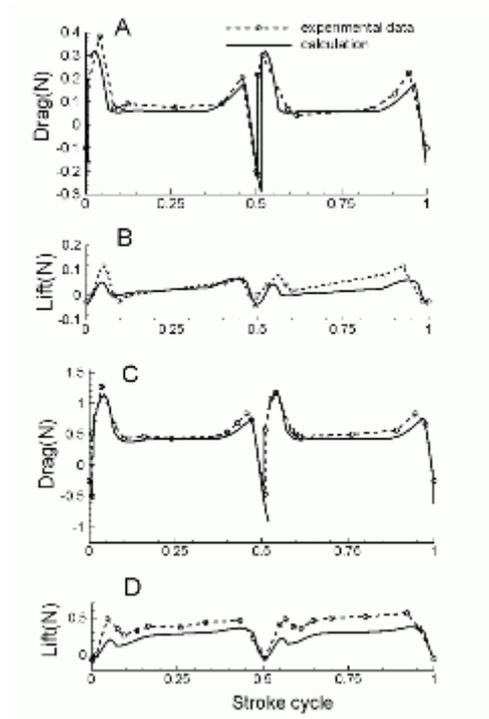

Fig.4

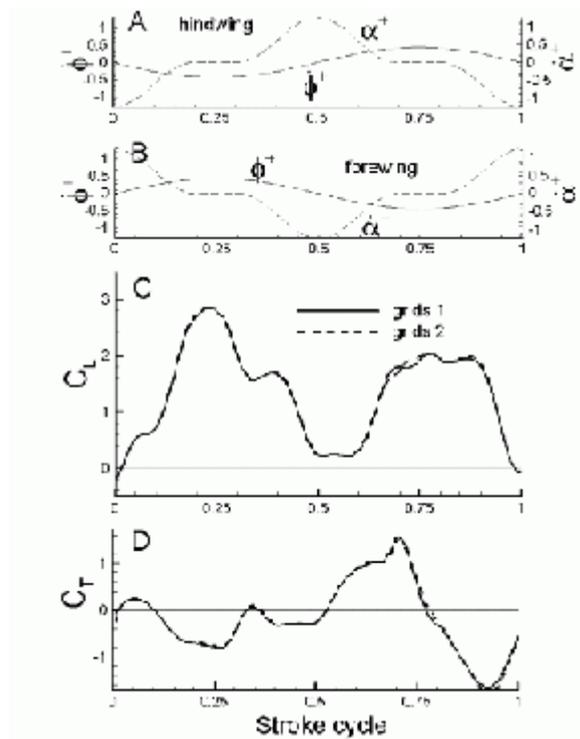

Fig.5

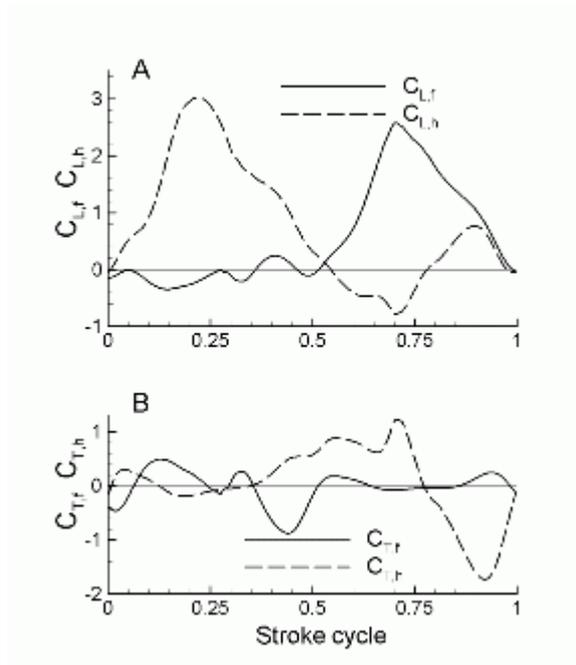

Fig.6

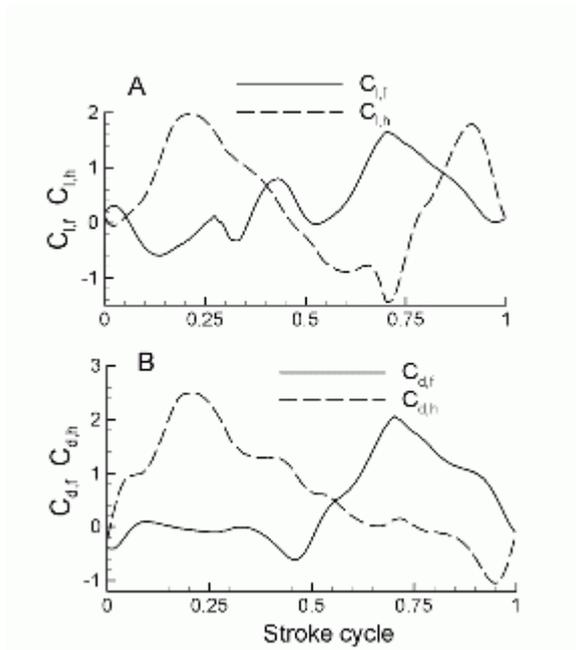

Fig.7

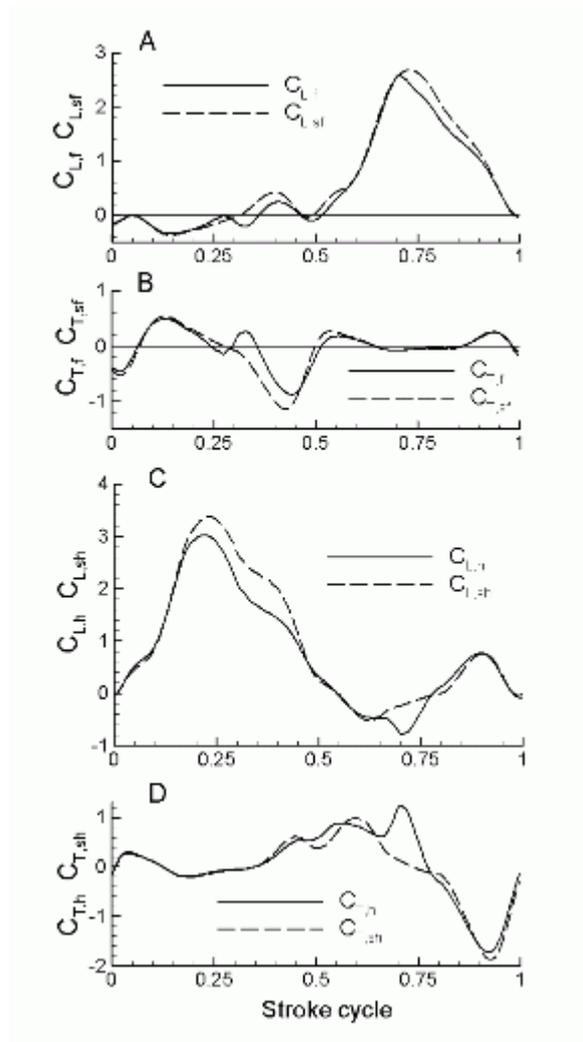

Fig.8

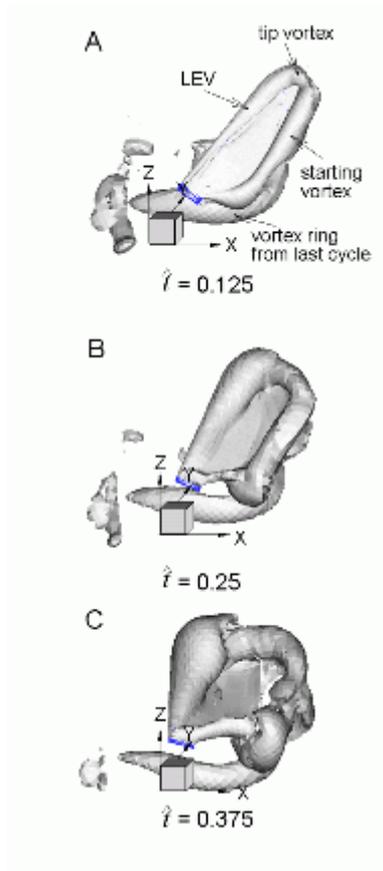

Fig.9(1)

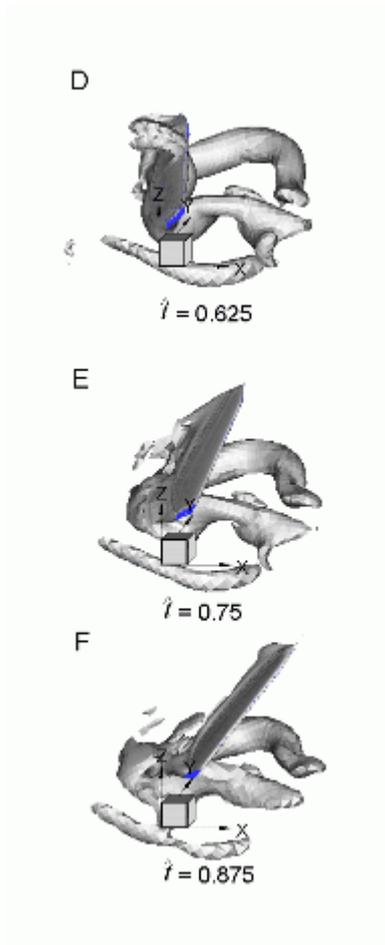

Fig.9(2)

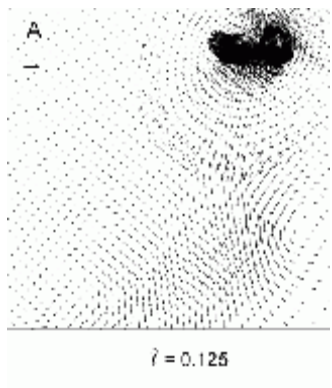

Fig.10A

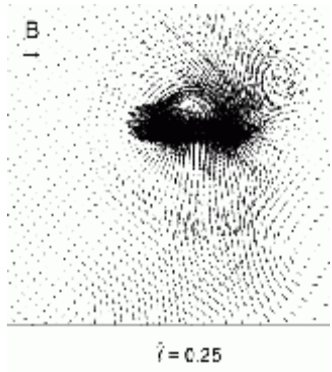

Fig.10B

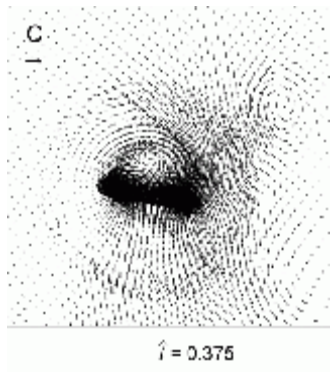

Fig.10C

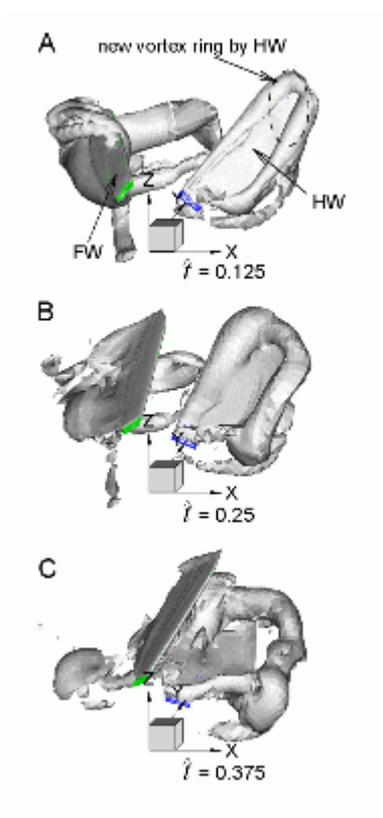

Fig.11(1)

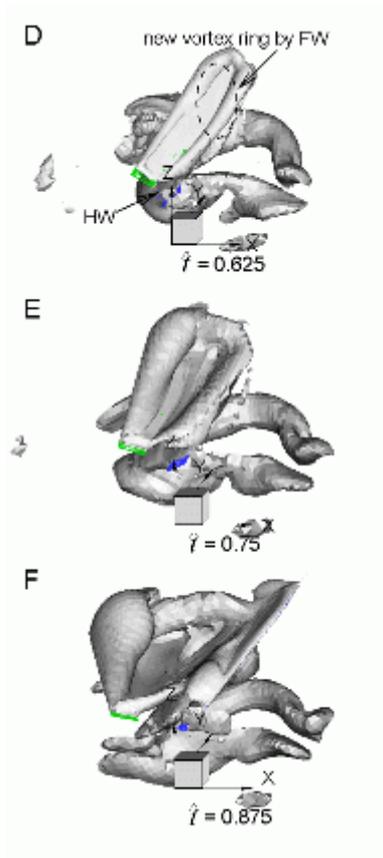

Fig.11(2)

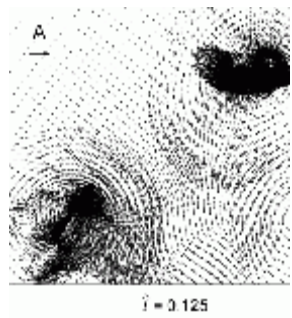

Fig.12A

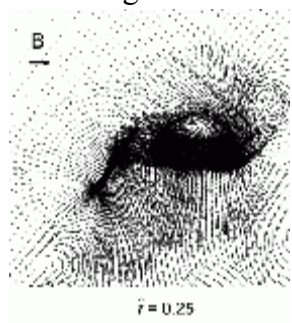

Fig.12B

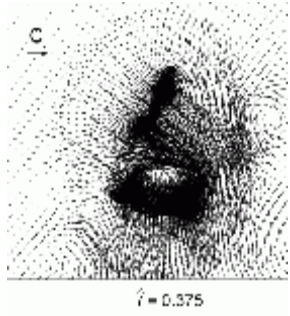
Fig.12C

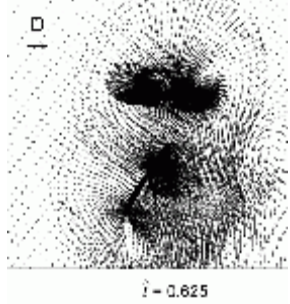
Fig.12D

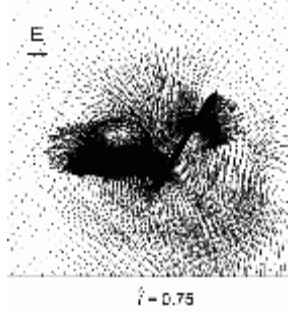
Fig.12E

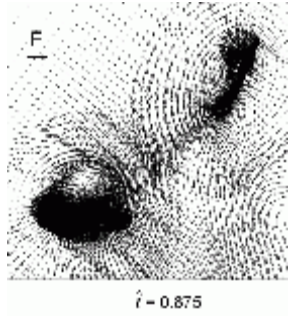
Fig.12F
31

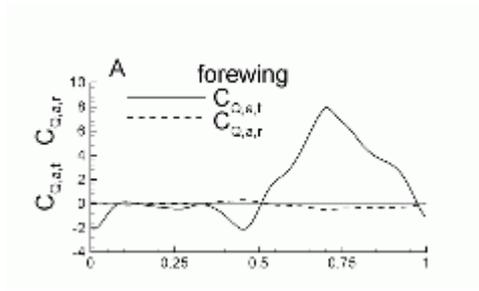

Fig.13A

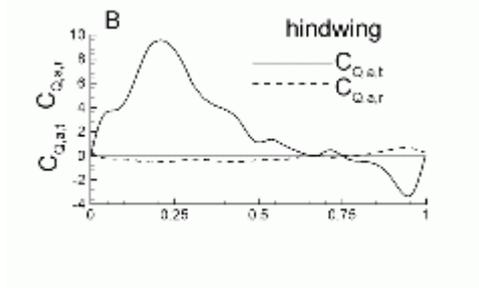

Fig.13B

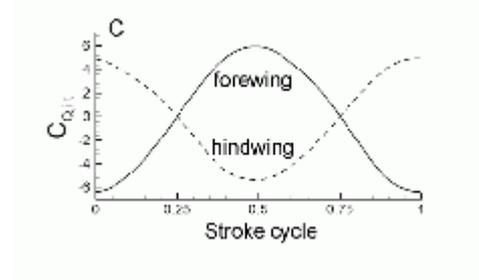

Fig.13C

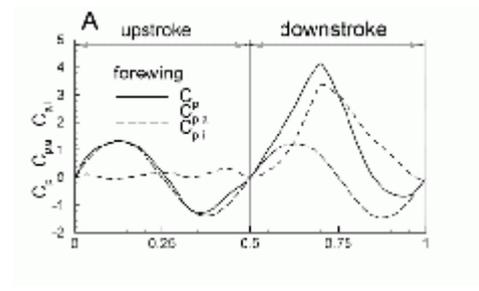

Fig.14A

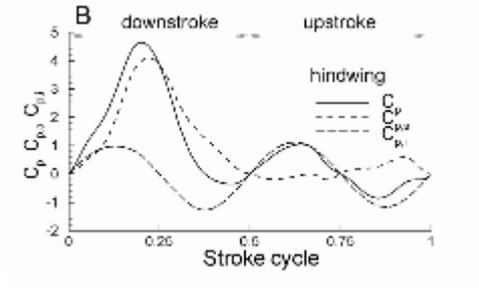

Fig.14B